%% file: MAIN.tex
\documentclass[lettersize,journal]{IEEEtran}
\usepackage{amsmath,amsfonts}
\usepackage{algorithm}
\usepackage{array}
\usepackage[caption=false,font=normalsize,labelfont=sf,textfont=sf]{subfig}
\usepackage{textcomp}
\usepackage{stfloats}
\usepackage{url}
\usepackage{verbatim}
\usepackage{graphicx}
\usepackage{cite}
\usepackage{lipsum}

\usepackage{textcomp}
\usepackage{xcolor}
\usepackage{threeparttable}
\usepackage{booktabs}
\usepackage{siunitx}
\usepackage{tabularx}
\usepackage{ragged2e}
\usepackage{subcaption}
\usepackage[noend]{algpseudocode}
\usepackage{multirow}
\usepackage{makecell}
\usepackage{hyperref}
\usepackage{inputenc}
\usepackage{shortcuts}
\usepackage{pgffor}
\usepackage{wasysym}
\usepackage{adjustbox}
\usepackage{tikz}
\usepackage{listings}
\usepackage{svg}
\usepackage{tcolorbox}
\usepackage{amssymb}

\usepackage[T1]{fontenc}

\newtcolorbox{mybox}[1][]{
  colback=blue!5!white,   
  colframe=blue!75!black, 
  fonttitle=\footnotesize,    
  fontupper=\footnotesize,
  coltitle=black,         
  title=#1,               
  boxrule=0.1mm,           
  arc=1mm,                
  auto outer arc,         
  width=\columnwidth,     
}

\lstdefinelanguage{JavaScript}{
  keywords={break, case, catch, continue, debugger, default, delete, do, else, finally, for, function, if, in, instanceof, new, return, switch, this, throw, try, typeof, var, void, while, with},
  morecomment=[l]{//},
  morecomment=[s]{/*}{*/},
  morestring=[b]',
  morestring=[b]",
  sensitive=true
}
\begin{document}

\title{Beyond the Scope: Security Testing of Permission Management in Team Workspace}



\makeatletter
\newcommand{\linebreakand}{%
  \end{@IEEEauthorhalign}
  \hfill\mbox{}\par
  \mbox{}\hfill\begin{@IEEEauthorhalign}
}
\makeatother

\author{\IEEEauthorblockN{Liuhuo Wan,}
\and
\IEEEauthorblockN{Chuan Yan,}
\and 
\IEEEauthorblockN{Mark Huasong Meng,}
\and
\IEEEauthorblockN{Kailong Wang,}
\and
\IEEEauthorblockN{Haoyu Wang,}
\and
\IEEEauthorblockN{Guangdong Bai\textsuperscript{}}\thanks{
\noindent\rule{0.9\columnwidth}{0.4pt} \\ 
\noindent
\textit{
\textbullet \quad L. Wan is with The University of Queensland, Australia.\\
\textbullet \quad C. Yan is with The University of Queensland, Australia.\\
\textbullet \quad M. H. Meng is with The Technical University of Munich, Germany.\\
\textbullet \quad K. Wang is with Huazhong University of Science and Technology, China.\\
\textbullet \quad H. Wang is with Huazhong University of Science and Technology, China.\\
\textbullet \quad G. Bai is with The University of Queensland, Australia and also with National University of Singapore, Singapore.\\
\textbullet \quad J. S. Dong is with National University of Singapore.\\
}},
\and 
\IEEEauthorblockN{Jin Song Dong}
}

\maketitle

\input{chapter/0_abs}

\input{chapter/1_intro}

\input{chapter/2_background}
\input{chapter/3_permission_management}

\input{chapter/4_attack_definition}

\input{chapter/5_overview}

\input{chapter/6_method}
\input{chapter/7_case_study}
\input{chapter/8_discussion}

\input{chapter/9_limitations}

\input{chapter/10_related_work}

\bibliographystyle{IEEEtran}
\bibliography{sample-base}

\end{document}

%% file: chapter/0_abs.tex
\begin{abstract}


Nowadays team workspaces are widely adopted for multi-user collaboration and digital resource management. 
To further broaden real-world applications, mainstream \bcp platforms, such as Google Workspace and Microsoft OneDrive, allow third-party applications (referred to as \textit{add-ons}) to be integrated into their workspaces, significantly extending the functionality of \bcp.
The powerful multi-user collaboration capabilities and integration of \textit{add-ons} make \bcp a central hub for managing shared resources and protecting them against unauthorized access.
Due to the collaboration features of \bcp, add-ons involved in collaborations may bypass the permission isolation enforced by the administrator, unlike in single-user permission management.

This paper aims to investigate the permission management landscape of \bcp add-ons. 
To this end, we perform an in-depth analysis of the enforced access control mechanism inherent in this ecosystem, considering both multi-user and cross-app features. We identify three potential security risks that can be exploited to cause permission escalation.  
We then systematically reveal the landscape of permission escalation risks in the current ecosystem. 
Specifically, we propose an automated tool, \toolname, to systematically test all possible interactions within this ecosystem.
Our evaluation reveals that permission escalation vulnerabilities are widespread in this ecosystem, with 41 interactions identified as problematic.
Our findings should raise an alert to both the \bcp platforms and third-party developers.

\end{abstract}

%% file: chapter/1_intro.tex
\section{Introduction}

\bigbcp offer a comprehensive toolkit to streamline business operations.
For example, users can manage the company's daily tasks through a spreadsheet. Other users can actively access this spreadsheet, assuming various roles such as viewer, commenter, or editor.
Prominent examples of \bcp include Google Workspace~\cite{Google_marketplace} and Microsoft OneDrive~\cite{microsoft2024apps}. The demand for online and remote collaborations, particularly exacerbated by the effects of the pandemic, has propelled the widespread adoption of these \bcp. 
Google Workspace, for example, boasts over three billion users~\cite{Google_workspace_user} and is adopted by nine million companies as their business solution. \bigbcp can be further enhanced by integrating third-party applications (named \addons) to supplement advanced features. 
These \addons integrate with external services such as Confluence, Evernote, GPT, Zoom, Dropbox, Webex, etc. 
While such integration enhances the user experience, it also introduces complexities in permission management when accessing sensitive resources hosted by \bcp. 
The lax management of \addon access control inevitably introduces new attack surfaces, posing risks to security-critical collaborations. Despite its significance, this issue has not been thoroughly studied before.

A \addon accesses resources managed by \bcp through host APIs, for tasks such as viewing documentation or replying to an email. 
These host APIs are regulated by different permissions.
The \addon employs a two-level permission management system, comprising OAuth permission scope checking and user role checking. 
In detail, during installation, the add-on requests OAuth tokens with permission scope, which are used to interact with resources managed by \bcp. 
For example, an \addon might request the OAuth permission scope to `\emph{view and edit all Google Documents}' from the user. 
However, some resources accessed by the \addon during runtime may not be owned by the user. 
If the user is assigned the viewer role and cannot edit the resource, the \addon installed by that user is similarly prohibited from modifying the resource. 

\begin{figure}[h]
  \centering  \includegraphics[width=1\columnwidth]{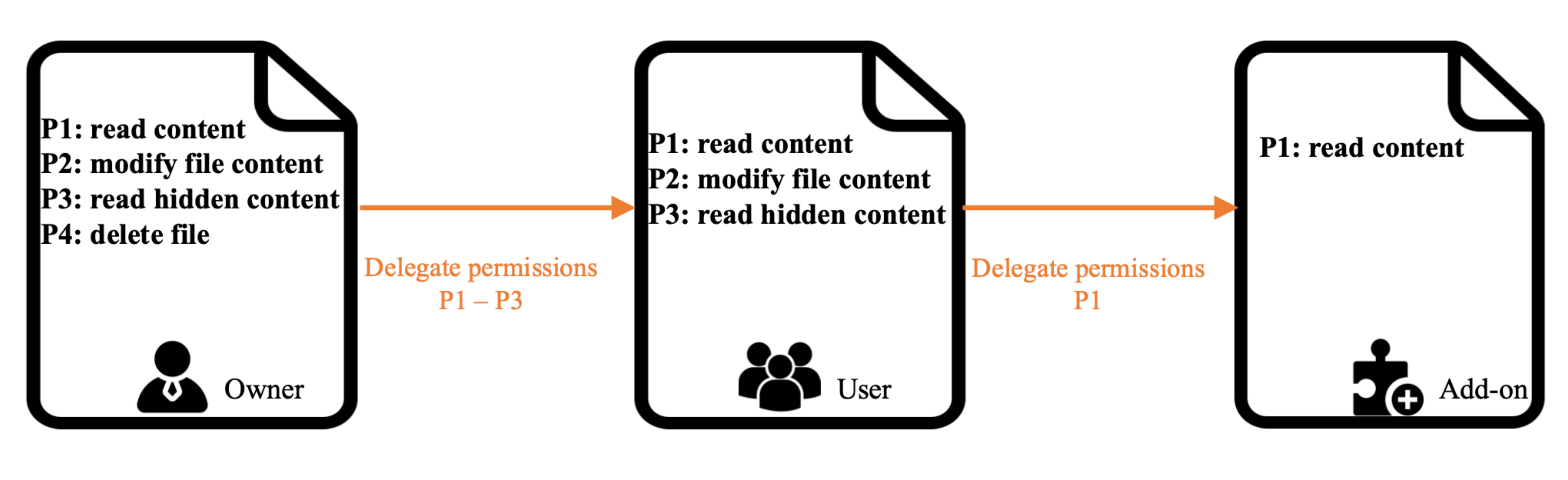}
  \caption{Permission delegation demonstration}
  \label{fig:permission_delegate}
\end{figure}

\textbf{Shadow of security risks.} 
\bigbcp supports multi-user collaboration through role-based permission management, where an owner retains full privileges and assigns partial privileges to other roles as shown in Figure~\ref{fig:permission_delegate}. It is expected that a user role does not exceed the permission scopes assigned by the owner. Similarly, \addons installed by specific a user role~(e.g., viewer), should adhere to the permission scopes delegated by this user.
This raises a critical question: does permission management remain consistent across different users and \addons. For example, \textit{can an unprivileged member install an \addon to access resources they are not entitled to, such as hidden rows or columns in a protected Spreadsheet, and even modify them?}
It is important to note that the installation of \addons in the private workspace (e.g., personal workspace) typically requires no vetting process~\cite{wan2024safe}. In such scenarios, if permission management is not well-established, users could potentially achieve permission escalation with the assistance of \addons.


Despite the potential security risks brought about by \addons, limited efforts~\cite{google_third_party_user_study,wan2024safe,bui2020xss} have been devoted to understanding the security issues they pose due to the black-box nature of both \bcp and \addons. 
Furthermore, the complexity of \bcp's features, including multi-user roles, complicates the analysis and testing of these security risks. 
Existing efforts~\cite{wan2024safe,bui2020xss,google_third_party_user_study} predominantly rely on manual interactions and inspections. 
Consequently, \textbf{there is a need for automated tools to test and analyze permission management considering the integration of \addons.}


To comprehensively evaluate the underlying permission escalation problem caused by the \addons, 
We systematically test and analyze the \addons' capability to utilize host APIs for achieving permission escalation.
Specifically, we develop a tool named \toolname to automatically generate test cases and evaluate the \addons' capability under different permission scopes. 
\toolname identifies 41 risky host APIs that \addon can exploit, leading to permission escalation. These APIs affect prominent services like Spreadsheets, Documents, Drive Files, and Forms. They allow unauthorized operations on resources, including viewing hidden data protected by the admin, unauthorized modifications of specific data groups, and interfering with existing members of shared resources.


\textbf{Contributions.} The contributions of this work are summarized as follows:
\begin{itemize}
     \item \textbf{An in-depth understanding of the permission system.} 
     We conduct a detailed analysis and reveal the complex permission control mechanisms within \bcp: the two-level permission checking for \addons. This enables us to understand the precise mechanisms that govern data access and permission requests. Based on this established knowledge, we can identify and formally define the permission escalation vulnerabilities in the current system. 

     \item \textbf{A systematic tool for the permission escalation testing.} To comprehensively detect problematic \addons' usage of host APIs that could lead to permission escalation, we have devised and implemented an automated tool called \toolname. 
     \toolname is built on a pipeline that includes permission categorization, test case generation, and user-role-based testing.
     Using \toolname, we can detect discrepancies during the permission-checking process.
     This systematic scanning approach enables us to identify all risky API usages within the ecosystem.

     \item \textbf{The landscape of permission escalation issues.}
     Our research has uncovered a large amount of problematic APIs. 
     Through detailed discussions supported by concrete case studies, we have identified potential attack scenarios that could arise from problematic API usage.
     Furthermore, we have proposed countermeasures to mitigate the permission escalation issues for \bcp and end-users.
\end{itemize}

\noindent\textbf{Ethics and Disclosure.} All our experiments are conducted using test accounts. The workspace is controlled, with the authors as the only members. The malicious \addons designed are only installed in the controlled \bcp and access limited resources. We did not distribute these malicious \addons into other \bcp or the public marketplace. All our attacks would not affect users and resources other than the authors' testing accounts. We ethically disclosed our findings to \bcp and received acknowledgment.

%% file: chapter/2_background.tex
\section{PROBLEM FORMULATION}\label{sec:bacground}


\subsection{Background and the Permission system}\label{subsec:permission-system}
Before the launch of \bcp, collaboration in organizations often relied on a combination of tools that, were not as seamlessly integrated or cloud-based as modern solutions. 
Different applications often require manual syncing and are not always designed for real-time, cloud-based collaboration. For example, Dropbox for file storage or Slack for messaging, operate in separate environments, requiring users to juggle multiple applications and interfaces. This fragmentation make it challenging to maintain a cohesive workflow, as data has to be manually transferred, and users have to manage different accounts and logins. \bigbcp bring a more integrated, cloud-native approach by combining email, chat, file storage, document editing, and scheduling tools into a single, unified platform. This integration allows for real-time collaboration on documents, seamless communication via email, and centralized file storage and sharing—all accessible through a web browser without downloading separate desktop applications. 
As a result, \bcp significantly simplify workflows, reduce the reliance on disparate tools and create a more streamlined, efficient collaboration environment for many organizations.

Sharing is one of the most powerful features supported by \bcp. Users can seamlessly share their resources and engage in real-time collaborative file editing, eliminating the need for redundant resource distribution. 

%% file: chapter/3_permission_management.tex
\noindent\textbf{Users.} 
There are four types of collaborators supported in the current ecosystem of \bcp, owner, editor, commenter and viewer. The owner has all privileges related to shared resources, while other roles have varying permission scopes, as suggested by their names. 
The owner can assign different roles to the collaborators.

\noindent\textbf{Add-ons.} The users can integrate \addons into their \bcp. Beyond interactions with legitimate users, \addons can access resources through user delegation as shown in Figure~\ref{fig:permission_delegate}. With permission scopes granted by users, \addons can access and manipulate resources stored within the user's workspace. These resources fall into two categories: those for which this user is the owner and those for which this user is a collaborator~(e.g., viewer or editor).

\subsection{Permission System}

\noindent \textbf{Objects}. 
There is a list of resources supported by the \bcp, such as documents, spreadsheets, presentations, and forms. 
Each resource exhibits a hierarchical structure composed of finer-grained components. For example, a spreadsheet consists of rows, each of which contains multiple cells. Every cell may carry its own set of attributes, such as content value, background color, and formatting. 
We refer to all of these components, at any level of granularity, as objects.

\noindent \textbf{Subjects}. 
In total, \bcp defines three types of subjects: (1) the \textbf{owner} of the object, (2) \textbf{collaborators} on the object (may be viewer, commenter or editor), and (3) \textbf{\addons} that can access predefined objects.

\noindent \textbf{Permissions}. 
We represent each permission in \bcp using a tuple $(subject, operation, object)$, which indicates who (the subject) can perform what action (the operation) on which resource (the object). 
The \bcp support five types of operation: \emph{create, view, comment, modify}, and \emph{delete}. 
These permissions may be applied at varying levels of granularity, ranging from an entire \texttt{Document} to nested elements such as a \texttt{Footnote} within that \texttt{Document}. For example, the permissions $(\textit{add-on}, \textit{delete}, \texttt{Document})$ and $(\textit{add-on}, \textit{delete}, \texttt{Footnote})$ indicate the capability of an \addon to delete the entire \texttt{Document} or only the \texttt{Footnote}, respectively.

\begin{figure}[t]
  \centering  \includegraphics[width=1\columnwidth]{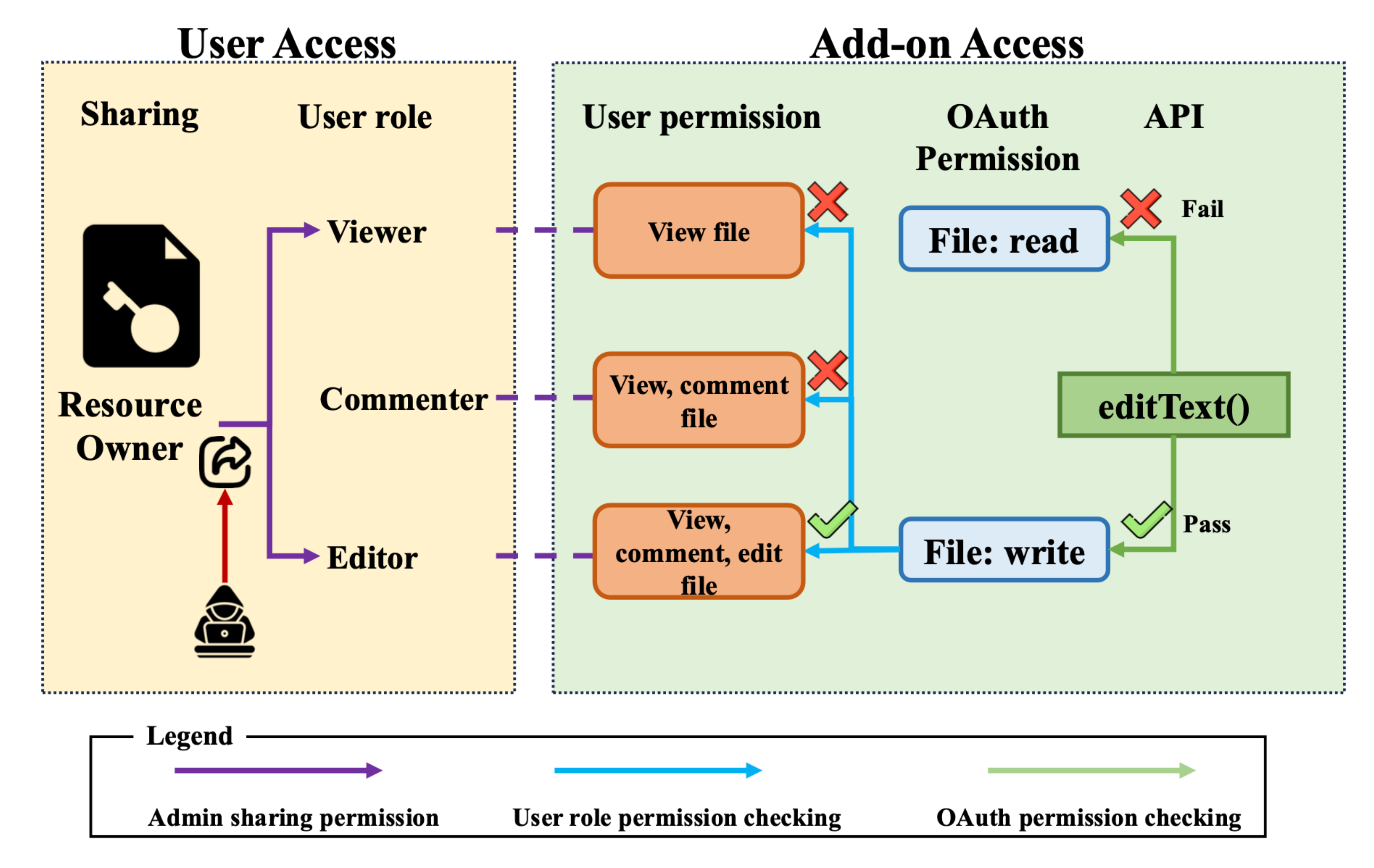}
  \caption{Access control mechanisms in the current ecosystem}
  \label{fig:permission_sys}
\end{figure}

Team workspaces enforce a two-level access control model for \addons, as illustrated in Figure~\ref{fig:permission_sys}. 
When the \addon utilizes host APIs like \texttt{editText()} to access the installer's resources, \bcp verify the authorized \emph{\textbf{OAuth}} permission scopes granted by the installer and the installer's \emph{\textbf{role}}~(collaborator's permissions) associated with the resource.

\paragraph{\textbf{Level 1: OAuth permission scope checking}} 
As illustrated by the green lines in Figure~\ref{fig:permission_sys}, when the \addon tries to execute the API \texttt{editText()}, \bcp first check the already authorized OAuth permission scopes by its installer. If the \addon has been granted \texttt{file:write} permission, it is allowed to write to the file and successfully passes the first-level permission checking. In contrast, if the \addon is only authorized with \texttt{file:read} permission, it will fail to execute the \texttt{editText()} API due to insufficient permissions.

\paragraph{\textbf{Level 2: User role permission checking}} 
After successfully completing the first-level permission check, \bcp then evaluate the installer's role on the resource. As shown by the blue lines in Figure~\ref{fig:permission_sys}, for the \texttt{editText()} API, even if the installer has granted the add-on with \texttt{file:write} permission, the second-level permission check will fail if the installer’s role for the resource is set to \emph{viewer} or \emph{commenter} without editing privileges. 
As a result, \bcp returns an exception.

Based on this two-level permission checking, we can conclude that the \addon is constrained by both granted permission scope and installers' roles: \permaddon $\subseteq$ \permauth $\subseteq$ \permuser. For example, when a user assumes the role of a viewer, both the actual user and the installed \addon are restricted from commenting on or editing this resource.

%% file: chapter/4_attack_definition.tex
\subsection{Permission Escalation Definitions}\label{subsec:attack_formalization}
Add-ons must adhere to the permission system enforced by \bcp. 
If \bcps fail to enforce these permission checks, \addons would achieve permission escalation, as indicated by the red line or cross mark in Figure~\ref{fig:permission_sys}. 
Our work focuses on \textbf{permission escalation} issues that arise from the add-on's use of host APIs.
We formalize our attack model as follows.

\begin{definition}[Permission control consistency]
In our definition, we let $\mathcal{O}$ represent the set of object fields~(e.g., spreadsheet, document); $\mathcal{A}$ denotes a set of operations that can be applied to the object; $\mathcal{S} = \{O, E, C, V\}$ represents the five user roles, where $O$ is short for $O$wner, $E$ is short for $E$ditor, $C$ for $C$ommenter and $V$ for $V$iewer. 

\noindent \textbf{User permission representation.} The user's permission set can be represented as a set of tuples \permuser $ = \{p_{u} \mid p_{u}:(s_{u}, a_{u}, o_{u}), s_{u} \in \mathcal{S}, a_{u} \in \mathcal{A}, o_{u} \in \mathcal{O} \}$. Each tuple  $p_{u}$ refers to when the user assumes the subject role $s_{u}$, this user can operate action $a_{u}$ on object $o_{u}$. 
    
\noindent \textbf{Add-on permission representation.} The \addon's permissions can be represented as a set of API permission scopes: \permaddon $ = \{p_{a}|p_{a}:(a, o), a \in \mathcal{A}, o \in \mathcal{O}\}$. The permission to execute each API $p_{a}$ is $(action, object)$. For example, to successfully execute the API \texttt{Document.editText()}, the add-on must acquire permission to \textit{edit}~(action) the \textit{text}~(object). 
    
\noindent \textbf{Permission consistency representation.} The two-level permission checking model can be represented as: $\mathbb{P}_{\mathrm{addon}}^{(U)} \subseteq$ (\permuser $\land Auth$), where $U \subseteq \mathcal{S}$ represents the set of role the user assumes on resources. For each \addon installed by user $U$, the \addon permission scope is confined by user role $U$ and cannot exceed the permission scope authorized by $U$, represented as $Auth$. In theory, we have \permaddon  $\subseteq$ \permauth $\subseteq$ \permuser  $\subseteq$ \permowner. 

\end{definition}

\begin{definition}[Permission escalation]
We delineate three types of permission escalation scenarios. The first two scenarios stem from two-level permission checking of \addons. The final scenario arises from resource-sharing permission checks. 
These three attacks are demonstrated as red crosses or red lines, as shown in Figure~\ref{fig:permission_sys}.
Each of these scenarios can lead to permission escalation~\cite{meng2023post,zhou2022uncovering}.

    \item \textbf{OAuth permission escalation~($E_{1}$):} (\permaddon $\not \subseteq$ \permauth), when the \addon installed on the user \bcp can exceed the authorized permission scopes, this violates the first level permission checking and brings in permission escalation. For example, if the \addon is authorized with the \texttt{file:read} permission but is still able to use \texttt{editText()} to write to the file (green arrows in Figure~\ref{fig:permission_sys}). This results in a permission escalation.
    
    \item \textbf{User role permission escalation~($E_2$):} (\permaddon $\subseteq$ \permauth) $\land$ (\permaddon $\not \subseteq$ \permuser), when an \addon installed on the user's \bcp can perform actions that the user is not permitted to take, it clearly violates second-level permission checks. For example, we find that:
    
    \begin{align*}
\left\{\, 
    p = (\mathit{view}, o) \,\middle|\,
    p \in \mathbb{P}_{\mathrm{addon}}^{(U)} \land 
    p \notin \mathbb{P}_{\mathrm{viewer}} 
\,\right\}
\end{align*}

    The $P_{\text{addon}}^{(U)}$ refer to add-on installed by user with role U.
    The \addon can read some specific content of the resource whereas its installer cannot (blue arrows in Figure~\ref{fig:permission_sys}). 
    
    \item \textbf{Sharing permission escalation~($E_3$):} 
   Besides permission checking when a user is assigned a role, we also consider permission sharing. As shown in Figure~\ref{fig:permission_sys}, sharing permissions are managed by the administrator, who can grant resource access to collaborators.
   However, if a user's assigned role is stealthily modified by \addons, altering the sharing topology configuration \config (purple lines in Figure~\ref{fig:permission_sys}) without administrator action, this also constitutes a permission escalation.
   We define this attack scenario as:

    \begin{align*}
    \mathit{Addon}.\mathit{modify}(\mathbb{C}) \rightarrow \mathbb{C}^{'}, \quad \text{where } \mathbb{C} \neq \mathbb{C}^{'}
     \end{align*}


\end{definition}

\noindent \textbf{Overview across representative platforms.} We present an overview of the two most popular and representative \bcp, Google Workspace and OneDrive. 
As shown in Table~\ref{tab:platform_overvew}, all of these platforms adopt a two-level permission check system. They enable user-sharing features with four types of user roles: owner, viewer, commenter, and editor. Both platforms are vulnerable to the first two permission escalation issues. However, OneDrive is robust against $E_{3}$ because it prohibits the \addons capability from modifying user roles.

\input{table/table-platform-overview}

%% file: table/table-platform-overview.tex
\begin{table}[t]
        \centering
        \scriptsize
        \caption{The implementation of \bcp and their potential permission escalation}
         \label{tab:platform_overvew}
        
        \begin{tabular}{l|c|c|c|c}

        \hline
        \textbf{Platform}                    &  Two-level permission system & $E_{1}$ & $E_{2}$ & $E_{3}$ \\ \hline
        \textbf{Google Workspace}            &  \cmark   & \redcircle{}  & \redcircle{} & \redcircle{} \\ \hline
        \textbf{Microsoft OneDrive}          &  \cmark   & \redcircle{}  & \redcircle{} & \blackcircle{}\\ \hline
        \end{tabular}

        \begin{tablenotes}
    \item  A check mark (\cmark) indicates that the platform adopts the two-level permission system identified in our study. A red circle (\redcircle{}) signifies potential vulnerability to permission escalation. A black circle (\blackcircle{}) signifies no vulnerability to permission escalation.
    \end{tablenotes} 
        
\end{table}

%% file: chapter/5_overview.tex
\section{Overview and Challenges} 
Detecting the predefined three scenarios requires a thorough understanding of the permission scope for each role in the current \bcp. However, the existing documentation on permission scope is incomplete. While previous work~\cite{wan2024safe} provides partial insights into permission scopes from the user side, it does not address the scope from the \addon perspective. In this study, we systematically investigate the relevant permission scopes in different perspectives (real user vs. delegated \addon). 


\paragraph{\textbf{Real-users}} To understand the proper permission scopes from user side, we establish five test accounts, each representing one of the five user roles in \bcp. Subsequently, we employ a manual approach to comprehensively explore real-user roles and their associated permission scopes by assuming different roles. 
The exploration of the user permission scope was a one-time effort and was conducted manually. 
We investigate the capabilities of each user role in performing the fundamental permission groups, which will be detailed in Section~\ref{subsec:permission_categorization}.

\paragraph{\textbf{Add-ons}} Unlike assessing the permission scope for real users, which can be evaluated through manual review by assuming different user roles, the permission checking of the \addon involve the invocation of a large number of host APIs. The security implications of these APIs cannot be effectively analyzed without the assistance of an automated tool. So we have designed an automation tool called \toolname that is capable of analyzing APIs, generating test cases, and automatically invoking APIs based on the official documentation~\cite{Gmail_sendMessage} provided by \bcp.

Our framework, called \toolname, a \textbf{\underline{T}}eam workspace \textbf{\underline{A}}dd-on AP\textbf{\underline{I}} testing tool, aims to test each API based on the official documentation provided by \bcp. It first categorizes each API to the corresponding permission group,  generates the correct test cases for each API,  then execute the API under different user roles and record the execution status. Finally, we check whether the three predefined permission escalation scenarios happened. 
For example, if a document editing API is tested within a viewer workspace and still executes successfully, it indicates permission escalation \textbf{$E_2$}. We identified three challenges when implemented the \toolname due to features specific to team workspaces.

\begin{itemize}
    \item \textbf{\emph{Challenge \#1: API hierarchy.}} 
    There is a hierarchical structure among host APIs as shown in Figure~\ref{fig:dep_call}, different from other web-based applications that use a flat HTTP request API for resource access~\cite{hazard_team_chat,bcp_permission_analysis}. Existing work~\cite{hazard_team_chat, bcp_permission_analysis, wan2024safe, wan2024bite} on OAuth-based web applications like Slack and GitHub cannot handle this API hierarchy. In particular, host APIs have interdependent relationships, where certain APIs must be invoked before others. For example, to execute an API that edits document content, the API call chain that returns the current active document, \texttt{DocumentApp.getActiveDoc()}, must be invoked first. Therefore, the primary challenge lies in precisely capturing the hierarchical structure of host APIs. 

   \item \textbf{\emph{Challenge \#2: Parameter generation.}}
   Certain APIs require context-dependent parameters for successful execution. For instance, the API \texttt{DriveApp.getFolderById(ID)} needs a valid folder \textit{ID}. To properly invoke this API, our \toolname must be able to supply a valid folder ID as input.

    \item \textbf{\emph{Challenge \#3: Efficiency.}} Team workspaces encompass a huge number of APIs, and testing every single one is neither efficient nor imperative for our objectives. Our primary goal is to identify permission escalation, necessitating the timely filtration of unnecessary APIs. For example, if a viewer lacks the capability to execute the \texttt{Document.addFooter()} API, subsequently rendering all methods reliant on the \texttt{addFooter()} meaningless for testing~(e.g., \texttt{Document.addFooter().setText(`footer')}), these cases should be excluded. Consequently, \toolname is specifically designed to address this challenge, ensuring the timely filtering of unnecessary test cases. 
   
\end{itemize}

%% file: chapter/6_method.tex
\section{Design of \toolname}

\begin{figure}
    \centering
    \includegraphics[width=\columnwidth]{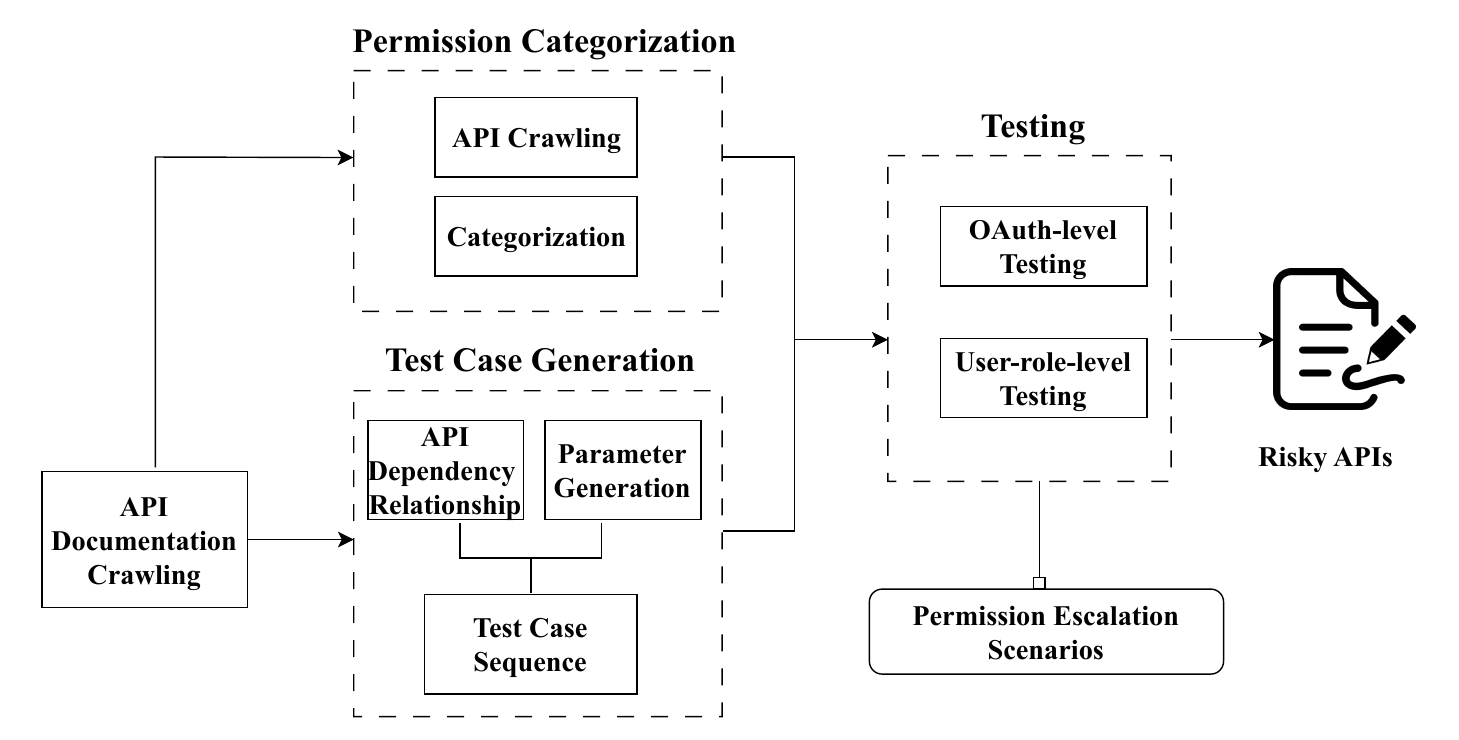}
    \caption{Workflow of \toolname} 
    \label{fig:overview}
\end{figure}

As shown in Figure~\ref{fig:overview}, \toolname consists of three major components: permission extraction~(Section~\ref{subsec:permission_categorization}), test case generation~(Section~\ref{subsec:test_case_generation}) and  testing~\ref{subsec:risky_api_testing}. Finally, we detail the risky APIs in Section~\ref{subsec:risky-api-iden}.


\subsection{API permission extraction} \label{subsec:permission_categorization}
\begin{figure*}
    \centering
    \includegraphics[width=1\textwidth]{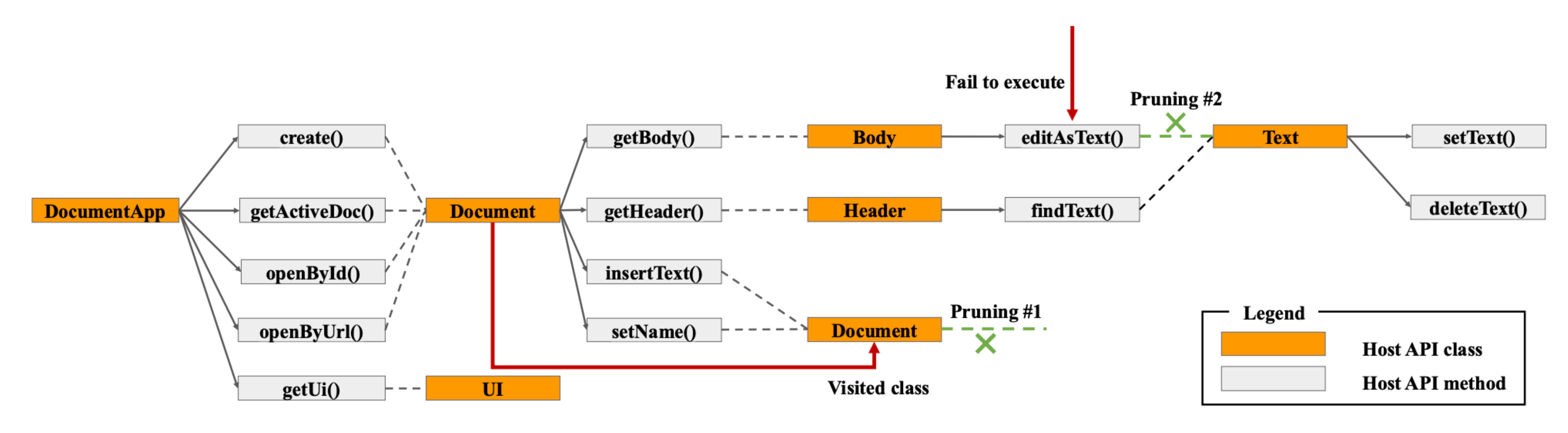}
    \caption{An example of dependency tree of the APIs} 
    \label{fig:dep_call}
\end{figure*}

We crawl all the API documentation including the API name, its parent class, description, parameters~(parameter name, type, description, etc.), tutorial code snippets~(if available), and the return type as shown in Figure~\ref{fig:code-snippets}, from the official developer website of \bcp~\footnote{Google developer reference: \url{https://developers.google.com/apps-script/reference}}. We use Selenium~\cite{Selenium}, the most widely used web crawling framework to automate this crawling process.  
As shown in Figure~\ref{fig:dep_call}, \bcp provides a standard hierarchy for objects at varying levels of granularity. For example, the objects \texttt{Body} and \texttt{Header} are fine-grained components of the \texttt{Document} object, while \texttt{Text} is a fine-grained component of the \texttt{Body} object. We adopt this hierarchy to derive the permission tuple for each API.
The total number of objects supported by each application is shown in Table~\ref{tab:fine_objects}.

We leverage the API's \textit{name}, \textit{description}, and the pre-extracted hierarchy to infer the intended semantic permission of each API.
We aim to categorize the extracted permission groups into one of the following types: $(create, object)$, $(view, object)$, $(comment, object)$, $(modify, object)$, $(delete, object)$ for later testing purpose.  
The \texttt{object} refers to the parent class of the API. For example, \texttt{Document.getBody()} would be categorized as $(\textit{view}, \texttt{Document})$.
In contrast, \texttt{Body.editAsText()} would be categorized as $(\textit{modify}, \texttt{Body})$.
We provide the GPT with a series of reasoning-based tutorials to enhance its chain-of-thought capabilities.
Notably, \bcp may employ diverse verbs to convey the same operation~(e.g., ``get a file'' and ``read a file'', both signifying the action of viewing a file). To address this, we adopt the powerful large language model - GPT capable of capturing the semantic meaning from diverse language descriptions. 
GPT is instructed to assign the correct label to each API with such format \textit{\addon, operation, object}. We use the gpt-4o-mini model~\cite{openai2024models} for this classification task. The example of the prompt we use is shown below.

\begin{mybox}
You are an engineer who would like to utilize the following API.

\textbf{\# Task Description}

I will provide you with the API name, its description and the context of object hierarchy. 

Your task is to categorize the API to one of the operation: (create, view, comment, modify or delete).

\textbf{\# Output Format}

[Output Format]
\end{mybox}

\begin{table}[]
    \centering
    \caption{Summary of object counts}
    \resizebox{\columnwidth}{!}{
    \begin{tabular}{p{1.5cm}|ccccccc|c}
    \toprule
    \textbf{Application} & Calendar & Document & Drive & Form & Gmail & Spreadsheet & Slide & \textbf{Total} \\
    \midrule
    \textbf{Number of Objects} & 7 & 36 & 6 & 34 & 6 & 58 & 47 & \textbf{194} \\
    \bottomrule
    \end{tabular}
    }
    \label{tab:fine_objects}
\end{table}


        


        

The OAuth permissions required for each API execution are documented in the \bcp documentation. Currently, \bcp provides only two types of authorization: read-only access and full access. Therefore, there is no need to match the extracted API permissions with OAuth scopes. However, the specific permissions of a subject collaborator, represented as a tuple $(collaborator, operation, object)$, are unclear. To address this, two authors independently reviewed the extracted permissions. They used the roles of viewer, commentor, and editor to verify whether the extracted permissions were actually granted to the collaborators. 
In total, there are 194 objects and 5$\times$ 194 permissions to be checked. The verification process took approximately eight hours to complete.



 \begin{table}[t]
        \centering
        \caption{Example of the returned class type of different APIs}
         \label{tab:returned_class_type}
         \resizebox{\columnwidth}{!}{
        \begin{tabular}{l|l|l}

        \hline
        & \textbf{Example 1} & \textbf{Example 2} \\ \hline
        \textbf{API}                    & \texttt{Document.getFootnotes()}      & \texttt{Document.getFooter()}  \\ \hline
        \textbf{Returned type}    & \texttt{Footnote[]}                   & \texttt{FooterSection} \\ \hline
        \textbf{\toolname strategy}     & \texttt{Document.getFootnotes()[0]}   & \texttt{Document.getFooter()} \\ \hline
        \end{tabular}
        }
\end{table}

\input{table/algo_1}

\subsection{Test case generation} \label{subsec:test_case_generation}

\subsubsection{\textbf{API dependency graph construction}} We reserve the dependency between APIs~(i.e., \textbf{\emph{Challenge \#1}}) and construct the \graph based on three components: \textit{classes},  all \textit{methods}~(APIs) of each class,  and the \textit{return type }of each method~(can be another class or void). To be detailed, we connect the \textit{class} with its \textit{methods}~(bold arrow in Figure~\ref{fig:dep_call}) and \textit{method} with its \textit{return type}~(dash line in Figure~\ref{fig:dep_call}) to construct this \graph.  For example, in Figure~\ref{fig:dep_call}, \texttt{DocumentApp} has multiple methods: \texttt{create()}, \texttt{getActiveDoc()}, \texttt{openById()}, etc. These four methods return with the same type of class \texttt{Document}. Further \texttt{Document} has multiple methods: \texttt{getBody()} that return the body of \texttt{Document} and \texttt{insertText()} that helps to insert text into the \texttt{Document}.

\subsubsection{\textbf{Parameter dependency}}
\begin{figure}[t]
    \centering
    \includegraphics[width=1\columnwidth]{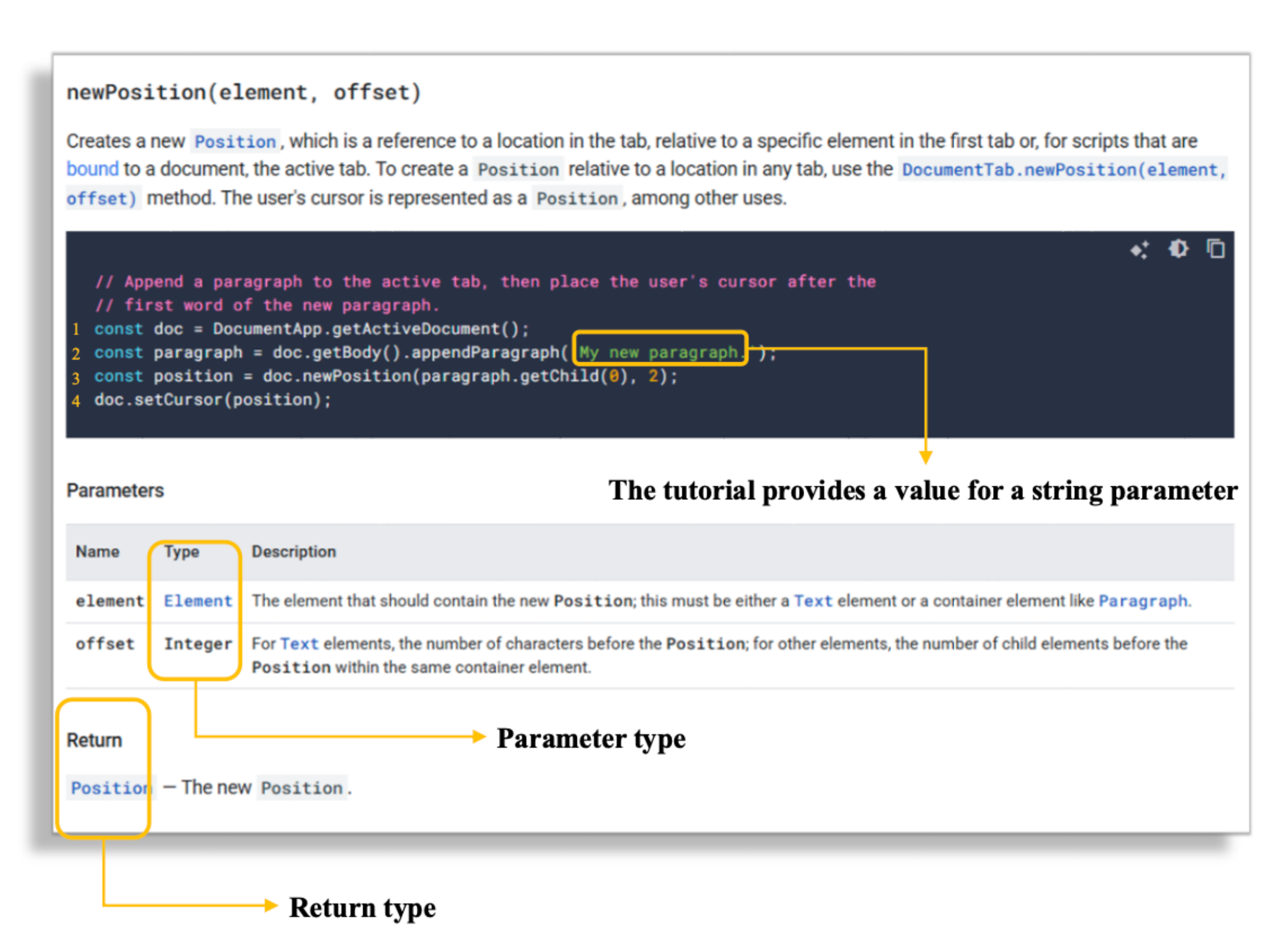}
    \caption{Tutorial code snippets provided by \bcp}
    \label{fig:code-snippets}
\end{figure}

Certain APIs require valid input parameters that are dependent on the outputs or responses of other APIs~(i.e., \textbf{\emph{Challenge \#2}}). For example, in Figure~\ref{fig:code-snippets}, the execution of \texttt{doc.setCursor(position)}~(line 4 in the code snippet) requires a valid value of \texttt{position}. We refer to such parameters as context-sensitive.
We categorize the API parameters into three types: context-sensitive parameters with tutorial snippets provided by \bcp, context-sensitive parameters without tutorial snippets, and non-context-sensitive parameters.

\noindent{\textbf{Tutorial available:}}
    We observe that \bcp provide detailed example code snippets for APIs requiring context related inputs, as shown in Figure~\ref{fig:code-snippets}. These code snippets are well-structured, context-sensitive, and generally do not require further modification. Consequently, we directly leverage these tutorial code snippets to capture the context parameter dependencies associated with such APIs. The only modification we apply to the tutorial code snippets is as follows: if the tutorial code snippets require \texttt{string} inputs to identify specific resources, we replace them with our customized string values, which will be detailed shortly.  

\noindent{\textbf{Tutorial unavailable:}}
    \begin{itemize}
        \item \textbf{Unique class:}
        If \bcps do not provide a code snippet for context-sensitive parameters, we search the \graph~(Figure~\ref{fig:dep_call}) to obtain a valid parameter. \bigbcp specify the \textit{type} of each required parameter in the official documentation, as shown in Figure~\ref{fig:code-snippets}. 
        For each parameter with a unique class type, as illustrated in Figure~\ref{fig:dep_call}, we traverse the connected \graph to find the shortest path that leads to the required \textbf{class}. 
        For example, \texttt{DocumentApp.getActiveDoc(). getPosition()} returns a valid input for the \texttt{Position} parameter.

        If multiple paths of the same length exist, we randomly select one. If the retrieved path returns an array of the required \textbf{class}, we fetch the first element from the array, as shown in Table~\ref{tab:returned_class_type}. Since our purpose is to test permission escalation rather than functionality, its actual value does not make a significant difference as long as the input is valid.

        \item \textbf{String constant:} 
        Specific APIs like \texttt{DocumentApp. openByName(``name'')} require a valid string constant for the \texttt{name} parameter. Unique \texttt{class} types can be precisely mapped through graph traversal whereas basic strings cannot. To address such cases, we maintain an attribute table of the accessed resources. This table records all the runtime values of string parameters~(e.g., id, url, name).
        

    \end{itemize}


\noindent \textbf{Non context sensitive:}
        For remaining parameters that do not fall into the previous categories, such as \texttt{Integer}, \texttt{Boolean} or context-irrelevant \texttt{String}~(line 3 in Figure~\ref{fig:code-snippets}), we simply enumerate several possible values. For \texttt{Integer}, we try values like 0, 1, and 10. For cases where multiple \texttt{Integer} inputs are required and have dependencies between them, we query the LLM model~(gpt-4o-mini) to fetch valid inputs. For example, in the API \texttt{copyFormatToRange(gridId, column, columnEnd, row, rowEnd)}, all four parameters are integers. According to the documentation of \bcp, \texttt{column} refers to ``the first column of the target range'', and \texttt{columnEnd} refers to ``the end column of the target range''. There is an implicit dependency where \texttt{columnEnd} should be larger than \texttt{column}. We leverage the LLM model, which is capable of handling such dependencies.

\begin{table}[t]
        \centering
        \scriptsize
        \caption{Example of the returned class type of different APIs}
         \label{tab:other_type_enumeration}
        
        \begin{tabular}{l|l}

        \hline
        Data Type & \textbf{Enumeration} \\ \hline
        \textbf{Integer}                    &  0, 1 , 5, 10  \\ \hline
        \textbf{Boolean}                    &  True, False  \\ \hline
        \textbf{Integer pair}               &  GPT-generated response \\ \hline
        \end{tabular}
        
\end{table}

\subsubsection{\textbf{Test case sequence}} 
The code listing illustrated in Figure~\ref{fig:code-snippets} demonstrates that the execution of the API \texttt{doc.setCursor(position)} (line 4) is contingent upon the successful execution of the call chain \texttt{doc = DocumentApp.getActveDocument()} (line 1), which is \textbf{\emph{Challenge \#1}}. 
In order to generate the correct sequence of test cases, we employ a breadth-first search (\textbf{BFS}) strategy combined with a pruning mechanism. As shown in Figure~\ref{fig:dep_call}, test case generation begins at the root class \textit{DocumentApp} and sequentially explores each method using BFS traversal. During traversal, when we encounter a previously visited class~(i.e., \textbf{\emph{Challenge \#3}}), we apply a pruning mechanism (referred to as \textbf{Pruning \#1}). For instance, the API \texttt{insertText()} in Figure~\ref{fig:dep_call} leads to a class \textit{Document} that has already been visited, we promptly prune this branch and discontinue generating test cases along this path. 
Algorithm~\ref{alg:code_api} shows the details of the \textbf{BFS} traversal from lines 3 - 11, and we apply the \textbf{Pruning \#1} in line 8.


Due to the implicit dependencies of resource operations, such as the prerequisite creation of a file before modification, and the necessity to test deletion as the final operation in a sequence. So, our test case generation would start with APIs within the create permission scope, followed by view, comment, and modify operations, and conclude with the delete permission scope. Similarly, for sharing permission, we adhere to such design: initiating the addition of collaborators first, followed by viewing, modifying, removing collaborators, or transferring ownership.


\subsection{Risky API testing}\label{subsec:risky_api_testing}
Due to the requirements of \bcp, the test cases can only be tested within the \addon configuration portal, as shown in Figure~\ref{fig:api-run-trigger}.

\subsubsection{\textbf{Development of add-ons}} Due to the intimate nature of add-ons, we designed and built a testing add-on that adheres to team workspace practices. We then distributed this add-on across workspaces with various user roles for testing purposes. It is worth noting that the design and development of the add-on is a one-time effort, taking approximately one hour to complete~\cite{Google_build_addon}. 
The \addon configuration portal is shown in Figure~\ref{fig:api-run-trigger}.

\subsubsection{\textbf{Test case execution}}
\begin{figure}[t]
    \centering
    \includegraphics[width=\columnwidth]{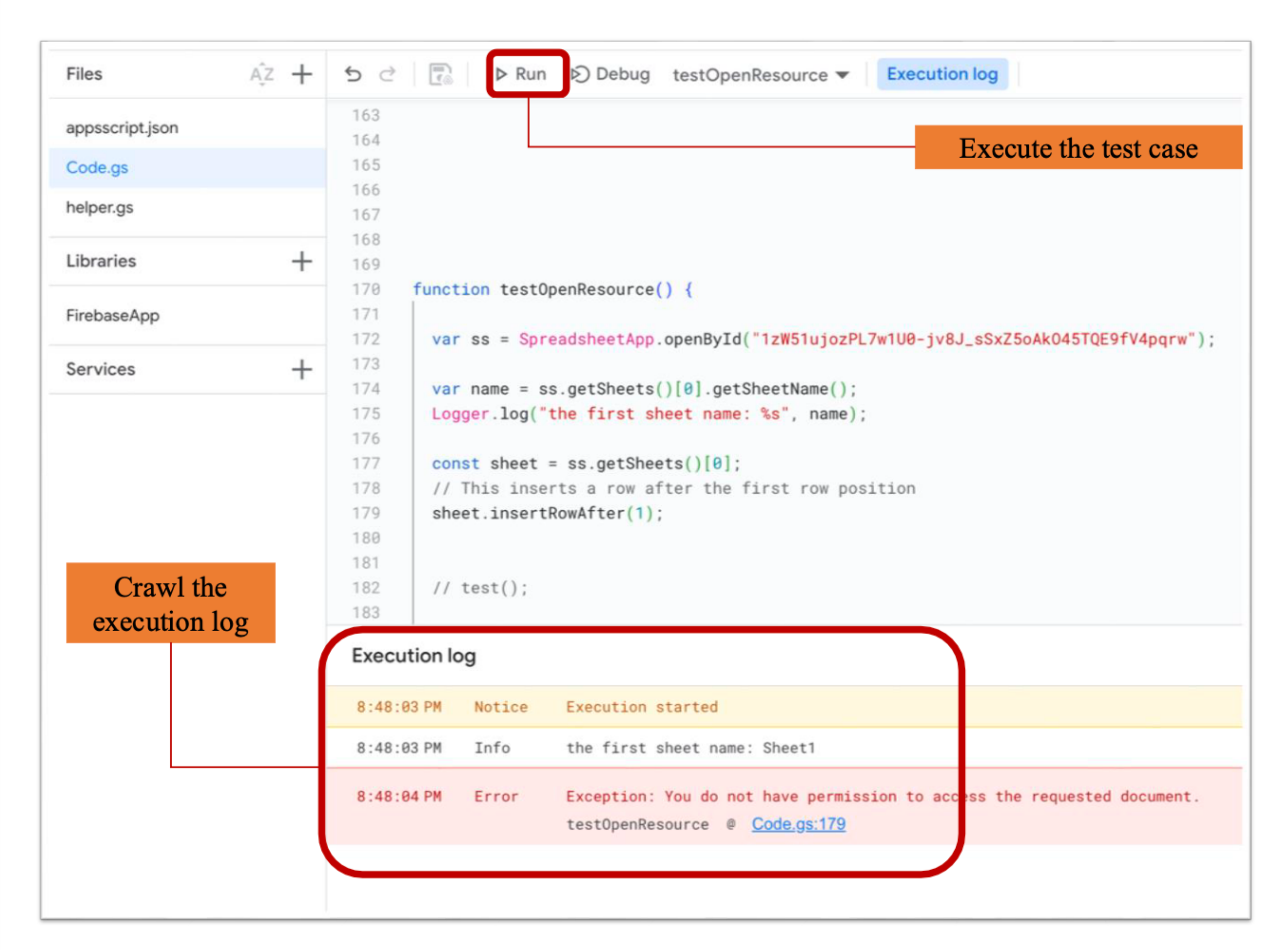}
    \caption{The API testing phase}
    \label{fig:api-run-trigger}
\end{figure}

At the start of each testing cycle, we fetch the candidate test cases~(to be tested) and update them into the project portal before execution. 
We use the Python library \textit{pyautogui}~\cite{PyAutoGUI_2025} to simulate various hotkeys and mimic developer behavior. 
This allows us to paste the testing candidate into the add-on configuration page (e.g., \texttt{Code.gs} in our scenario) using the \textit{Ctrl} + \textit{V} command. \textit{Pyautogui} also simulates the save action (\textit{Ctrl} + \textit{S}) to save the changes to the add-on project. 
After updating the test cases, we automated the click actions the \textit{Run} buttons using \textit{pyautogui} and recorded the execution log~(bottom part of Figure~\ref{fig:api-run-trigger}) of the test case as an indicator of its execution status.

\emph{OAuth-level testing.}
Based on the API permission categorization results, we utilize a progressive strategy when testing OAuth-level permission escalation~($E_{1}$). We begin by granting read-only permissions to the \addons and then test whether they can successfully execute APIs that belong to the edit or delete groups. Next, we extend the permissions to include read and edit scopes to see if the \addons can execute APIs that belong to the delete group. All testing was conducted using an owner account.

\emph{User-role-level testing.} We created five user profiles for testing. For each new round of testing, we utilized the same resource template to evaluate the execution of test cases. Each test case was tested under three different roles: viewer, commenter, and editor. 
Our goal is to test whether \addons can bypass user-role-level permission checks. So the installer pre-authorizes all required permission scopes, ensuring that \addons pass the OAuth-level check.

We monitor the execution status of each test case. 
If one API fails during testing~(Figure~\ref{fig:dep_call}), we proceed to prune all subsequent API calls~(\textbf{\emph{Challenge \#3}}) that depend on the successful execution of this API, referred to as \textbf{Pruning \#2}. If the API passes the testing, we document the successful execution of the API and then continue evaluating subsequent API calls that rely on this one.
The details of \textbf{Pruning \#2} are demonstrated in lines 12–18 of Algorithm~\ref{alg:code_api}.

\subsubsection{\textbf{API execution result monitor}} Although the response from a successfully invoked API may vary, the response from an invalid API invocation is more consistent. After conducting an in-depth review of the team workspace developer documentation~\cite{Gmail_sendMessage}, we found that when the API fails to execute, the \addon will return a specific error message, such as ``Exception: You do not have permission to access the requested document'' and stop execution immediately. 
Whenever the execution returns error messages, the execution log outputs the type \textit{Error}, as indicated in Figure~\ref{fig:api-run-trigger}. We mark type \textit{Error} as the indicator of a failed execution.
On the other hand, API executions that do not return any error messages are deemed successfully executed and pass the testing process.

\subsection{Risky API identification}\label{subsec:risky-api-iden}
\toolname evaluates whether the attack scenarios outlined in Section~\ref{subsec:attack_formalization} are plausible by examining the runtime results of API invocations under different user roles. For example, if an API invocation that exceeds the user's permissions completes successfully without returning any errors, it suggests a potential security violation. However, verifying whether APIs designed to retrieve information return valid responses~(as opposed to NULL or encrypted text) necessitates additional manual verification. 
We categorize the potentially risky APIs based on the criteria established in Section~\ref{subsec:attack_formalization}.

\noindent{\textbf{Detection of $E_{1}$:}} If the \addon is able to execute an API that falls outside the permission scope authorized by the user, it indicates a potential security issue or permission bypass: $p_{a}$ $\in$ \permaddon $\land$  $p_{a} \not\in$ \permauth, this will be detected as attack scenario \textbf{$E_1$}. 

\noindent{\textbf{Detection of $E_{2}$:}} Similarly, if the \addon is able to execute an API that operates on resources but the user cannot: $\{p_{a}: (a, o) \mid  p_{u}: (s_{u}, a, o), p_{a} \in$ \permaddon $\land p_{u} \not\in$ \permuser \}, this will be detected as \textbf{$E_2$}. 

\noindent{\textbf{Detection of $E_{3}$:}} 
Finally, if \addon executes an API that modifies the current sharing configuration \config of the resource without notifying the administrator, it will be classified as \textbf{$E_3$}.


\input{table/table_test_api}

\input{table/table5}

%% file: table/algo_1.tex
\begin{algorithm}[t]
\footnotesize
 \caption{Testing for API calls} \label{alg:code_api} 
 \begin{algorithmic}[1]
  \Statex{\textbf{Input} $G$: Knowledge graph}
  \Statex{\textbf{Output} $C$: candidates of testing $APIs$}
  \State $T \leftarrow \emptyset$, $Visited \leftarrow \emptyset$
  \State $T \leftarrow [CalendarApp$, $DocumentApp$, $DriveApp$, $FormApp$, $GmailApp$, $SlidesApp$, $SpreadsheetApp]$
  
  \Function{Generate\_test\_cases()}{}
  
  \While{$T \neq \emptyset$}
        \State $node = T.pop()$
        \For{all method $api$ of $node$ in $G$}
        \State $Class_{type}$ = $API_{returnType}$
        \If{$Class_{type}$ is not visited}   \Comment{\textbf{Pruning \#1}}
            \State $T.add(Class_{type})$
            \State $C.add(dep:node, api)$
            \State $Visited \leftarrow Visited \cup Class_{type}$
        \EndIf
        \EndFor
    \EndWhile
  \EndFunction
    
  \Function{Testing}{$C$}
  \For{each candidate $c$ in $C$}
        \State $response$ = Execute c
        \If{$response \in errors$}          \Comment{\textbf{Pruning \#2}}
            \State filtering all APIs depend on the execution of $c$
        \EndIf
        \If{$response \notin errors$}
            \State record the $response$
        \EndIf
  \EndFor
  \EndFunction
 \end{algorithmic}
\end{algorithm}

%% file: table/table_test_api.tex
\begin{table}
  \centering
  \scriptsize
  \caption{\toolname performance}  
  \label{tab:test_api}
  \begin{tabular}{ l | l | l | p{1.5cm} | l}
    \toprule
    \textbf{Host App} & \textbf{\# APIs} & \textbf{\# Tested APIs} & \textbf{Potential risky APIs} & \textbf{Risky APIs}  \\ \hline
    Calendar  & 218 & 132  &  0 & 0 \\ \hline
    Document       & 846 & 47 & 8 & 6   \\ \hline
    Drive     & 152 & 142 & 25 & 14 \\ \hline
    Gmail     & 167 & 101 & 0 & 0 \\ \hline
    Forms     & 418 & 84 &  2 & 2\\ \hline
    Spreadsheets    & 1784 & 624 & 15 &15  \\ \hline
    Slides    & 938 & 46 & 4 & 4 \\ 
    \bottomrule
  \end{tabular}
\end{table}

%% file: chapter/7_case_study.tex
\section{Evaluation}



\noindent \textbf{Scope.} Following the practice of previous study~\cite{wan2024safe,harkous2017if,google_third_party_user_study}, we implement \toolname and evaluate its performance on the most representative \bcp, Google Workspace, which occupies around 73.04\% market share~\cite{wan2024safe}. 

\subsection{Experiment setting}
We collected all official documentation, totaling \totapi APIs across seven host applications, with an average of about 646 APIs per host application. To initiate automated API testing, we created multiple testing accounts representing different user roles for collaboration. 
Our experiments were conducted on a series of computers assigned to different user roles to avoid any potential fingerprinting.

\subsection{\toolname performance}
The host application email has only one user role (owner) and does not support the collaboration feature, so we focus our testing on OAuth-level permission violations for email. For the remaining six host applications, we test all three permission escalations. The specific risky APIs for each host application are listed in Table~\ref{tab:test_api}.

\paragraph{\textbf{Effectiveness of API categorization}} We apply manual efforts to measure the effectiveness of \toolname.
Due to the uneven distribution of permission groups, random sampling would result in APIs belonging to the \textit{view} or \textit{modify} categories dominating the selection, which would not accurately reflect the performance of \toolname.
Following the methodology outlined in previous studies~\cite{hazard_team_chat,taintMini}, we randomly select around 200 APIs~(out of \totapi, 20 APIs per permission group) and check the correctness of permission categorization. 
Our manual inspection identified only two problematic cases, achieving an accuracy of 99\%. One such case involves the API \texttt{SlideApp.newAffineTransformBuilder()} which is categorized into \textit{modify slide} permission group. This API returns a new \texttt{AffineTransformBuilder} that assists developers in constructing an \texttt{AffineTransform}. However, this builder does not affect the corresponding Slides unless explicitly applied to a specific slide like \texttt{AffineTransform.insertToSlide(page)}. The method \texttt{waitForAllDataExecutionsCompletion( timeoutInSeconds)} also encounters a similar issue.


\paragraph{\textbf{Effectiveness of test case generation}}
We evaluate the performance of \toolname during the test case generation phase, with details provided in Table~\ref{tab:test_case_generation}. 
Our performance evaluation focuses only on APIs that are not filtered out. For APIs that are filtered out, generating test cases would be meaningless, as they will not be tested.

The majority of APIs are filtered through \textbf{Pruning \#1, \#2}. Out of the \testedapi APIs being tested, 482 APIs require no parameter input, while the remaining require the parameters. Among the 622 APIs that require a valid parameter, 481 have a tutorial code snippet available, and we directly apply these tutorial codes for our test case generation. For these 70 APIs requiring a valid \texttt{string} constant, we query our attribute table to fetch the corresponding string value. Around 38 APIs require other parameters like \texttt{Integer} inputs, or \texttt{Boolean} inputs.

Following the same methodology, we extracted 50 APIs from each category, for the categories \textit{unique class type} and \textit{others} that contain fewer than 50 APIs, we sampled all of them. This results in a total of 221 APIs. 
We manually evaluated the validity of \toolname-generated test cases by executing them and observing whether any runtime errors occurred. 
The precision of \toolname-generated test cases across different groups is summarized in Table~\ref{tab:test_case_generation}. Our evaluation shows that only a small proportion of the \toolname-generated test cases are invalid. For around 91\% APIs~(require no parameters or with tutorial), \toolname achieves excellent performance - 100\% accuracy. We investigate reasons for invalid test cases for the remaining three categories: \\
\noindent \textbf{Unique class type.} All four invalid test cases in Table~\ref{tab:test_case_generation} are introduced by the special \textit{Enum} parameter. For example, one API \texttt{Presentation.appendSlide(predefinedLayout)} requires the parameter of class type \texttt{predefinedLayout}. After reviewing the documentation of \texttt{predefinedLayout}, we find this is a special \textit{Enum} class defined by \bcp. To use it, the developer must be able to provide a predefined type like \texttt{predefinedLayout = SlidesApp.PredefinedLayout.BLANK}. 
However, the current design of \toolname cannot generate valid input for the \texttt{Enum} class, as these types are not included in the \graph.
\\
\noindent \textbf{String constant}. The two problematic cases in this category are caused by the implicit meaning of the \texttt{string} parameter. For example, the API \texttt{Spreadsheet.getAs(contentType)} requires a string parameter \texttt{contentType}, which is not included in our pre-built attribute table. Upon investigating the documentation, we found that \texttt{contentType} must follow a specific format, as shown below:

\begin{mybox}
\textbf{contentType:} the MIME type to convert to. For most blobs, \texttt{application/pdf} is the only valid option. For images in BMP, GIF, JPEG, or PNG format, any of \texttt{image/bmp}, \texttt{image/gif}, \texttt{image/jpeg}, or \texttt{image/png} are also valid. For a Google Docs document, \texttt{text/markdown} is also valid.
\end{mybox}
In this case, \toolname fails to provide a valid input for \texttt{contentType}. \\

\input{table/table_test_case_generation_performance}

\paragraph{\textbf{Effectiveness of API testing}} 
Although \bcp offer a vast array of APIs, we do not need to invoke all of them for our testing. 
We avoid testing scenarios that are inherently safe, such as an editor executing APIs requiring edit permissions under full authorization. Despite the large number of APIs, pruning strategies (\textbf{Pruning} \textbf{\#1}, \textbf{\#2}) effectively filter out 74\% of them as shown in Table~\ref{tab:test_api}. Out of the \totapi APIs, only \testedapi APIs would go through the risky API testing phase. The pruning strategies significantly improve the testing efficiency of \toolname by eliminating unnecessary paths.
The testing phase takes about four hours in total, with each API averaging 12 seconds to complete.


\subsection{Large-scale scanning and findings}

In Table~\ref{tab:test_api}, the last two columns present the number of potentially problematic APIs identified by \toolname. More specifically, \toolname reported 54 APIs that may align with our attack scenarios. Since some APIs return null values or encrypted text that do not disclose confidential information, we conducted a manual inspection and confirmed that 41 of them pose security risks and match our criterias. 
For example, while the API \texttt{File.getSharingPermission()} successfully executes, it returns a \texttt{NULL} value and does not expose confidential data that the user cannot access. 
These false positives were filtered out through manual review. 

Among the identified risky APIs, 17 pose $E_{2}$ risks, 21 pose $E_{3}$ risks, and none are classified as $E_{1}$ risks.
The output of \toolname shows that \bcp are robust against \textbf{OAuth-level} attack scenarios~($E_{1}$). Team workspaces enforce strict OAuth-level permission check, ensuring installed \addons do not exceed their authorized scopes.
This robustness is expected.
However, numerous permission escalations ($E_{2}$) arise due to inconsistent checking of the \textbf{resource user role} for both users and \addons. Furthermore, \addon's ability to modify \config  without the admin's awareness ($E_{3}$) puts all invited collaborators at risk.
We discuss the identified risky APIs that \toolname detects and demonstrate the attacks that can be launched through several case studies.



\paragraph{\textbf{Hidden value leakage}} 
\bigbcp provide rich functionality for spreadsheets, including the ability to hide certain sheets, rows or columns. Only collaborators with \emph{unhide} permissions can unhide and view these hidden values. However, \bcp impose no restrictions to \addon APIs, allowing viewers without \emph{unhide} permissions to recover hidden values via the exposed APIs. In our experiment, we concealed a \emph{salary} column and prevented viewers from copying, downloading, or printing the spreadsheet (settings that the owner can adjust as needed). Under this setting, viewers can only view the resource content online and cannot view or recover the hidden salary column. We first tested the viewer's ability to execute \texttt{row.unhide()} API  and it returned an error. However, by utilizing \texttt{row.getCell()} API, \addon installed on the viewer account can successfully fetch the hidden values even if the viewer user is forbidden from viewing the hidden values. 

\begin{figure}
    \centering
    \includegraphics[width=0.9\columnwidth]{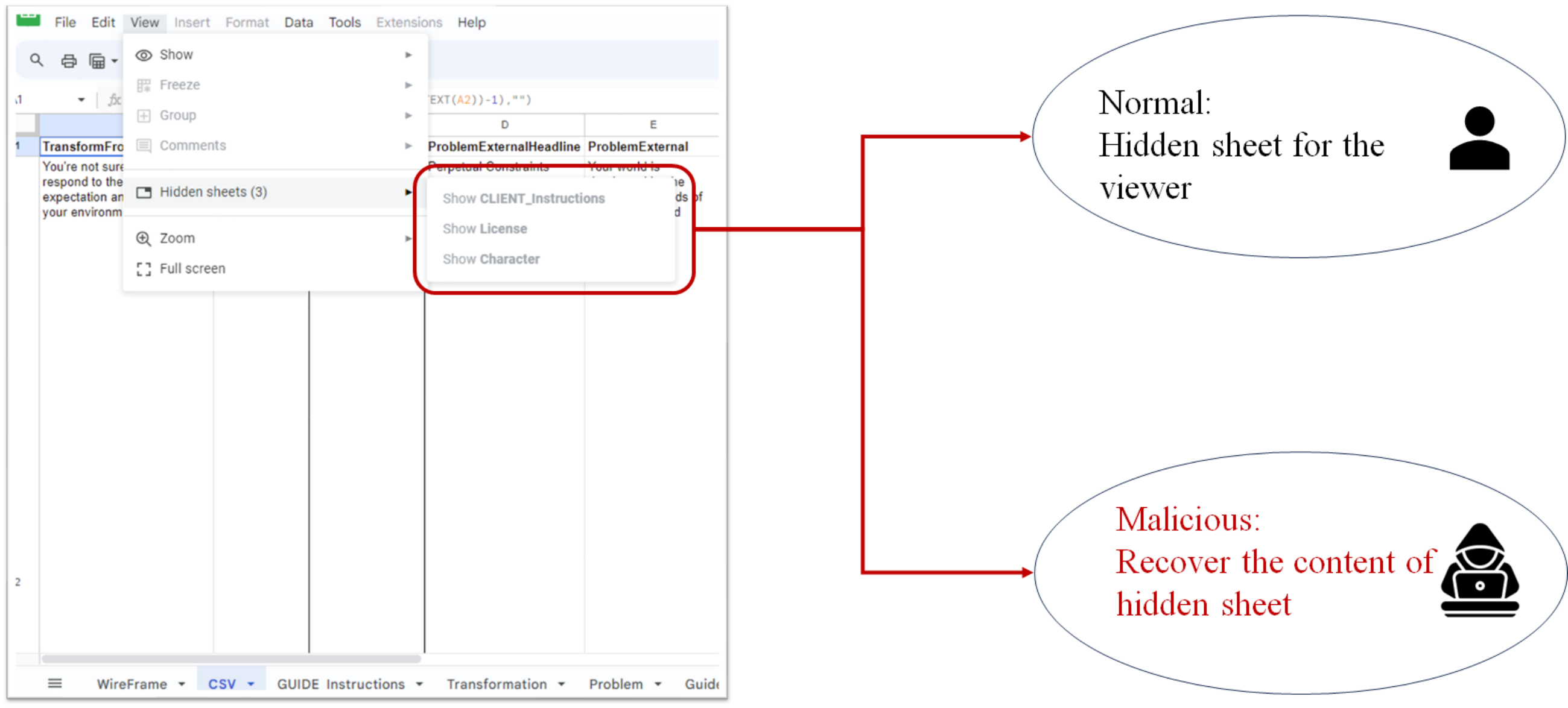}
    \caption{Attack: hidden sheet recovery}
    \label{fig:hidden-sheet_concealment}
\end{figure}

Even worse, team workspaces would explicitly prompt a warning notification when the owner hides a specific sheet.
\begin{mybox}
Use the \emph{View} menu to unhide sheets. All editors of this spreadsheet can view and unhide hidden sheets.
\end{mybox}
In this case, viewers are prohibited from accessing the hidden sheets; however, by utilizing the fine-grained API \texttt{Range.getCell()}, an attacker without proper permissions~(editor in this case) can recover the entire hidden sheet as shown in Figure~\ref{fig:hidden-sheet_concealment}.

\paragraph{\textbf{Protected range edition}}
Additionally, we observed that spreadsheets allow the owner to set specific ranges to be protected and only editable by a specific group of people (referred to as privileged editors), rather than all editors~(referred to as common editors). However, common editors are still permitted to edit the unprotected ranges. Once the protected range is \textit{grouped}, common editors are forbidden from editing or ungrouping the whole group. However, our experiment shows that in this scenario, even though common editors are unable to either ungroup the range or edit each cell in the group, APIs exposed to \addon can still pass the execution and edit the cell value in this group. Even though the \addon fail to execute the \texttt{range.ungroup()} API.


\paragraph{\textbf{Modification of user subject}}
The 21 APIs allow \addon to modify collaborators' subject roles without awareness of administrator.  
We observe that \bcp have already banned the use of the dangerous API~\cite{google2024APIConsent} \texttt{transferOwnership()}. Currently, transferring ownership must be done manually by the actual user rather than initiated by an \addon. To prevent misuse, \bcp should also implement stricter management of the APIs that could be exploited by \addons.

%% file: table/table_test_case_generation_performance.tex
\begin{table}
  \centering
  \scriptsize
  \caption{Test case generation }  
  \label{tab:test_case_generation}
  \begin{tabular}{ l | l | p{2.5cm} | l}
    \toprule
    \textbf{Category} & \textbf{\# APIs} & \textbf{\# Invalid test case (out of 50 sampled)} & \textbf{Precision}   \\ \hline
    \multicolumn{4}{c}{\textbf{No Parameter Required}} \\ \hline
     No parameter           & 518  & 0 & 100\%\\ \hline
    \multicolumn{4}{c}{\textbf{Tutorial Available}} \\ \hline
     Tutorial               & 517 & 0 &   100\% \\ \hline
    \multicolumn{4}{c}{\textbf{Tutorial Unavailable}} \\ \hline
    Path dependency         & 33 & 4  & 87.9\% \\ \hline
    String constant         & 70 & 2  & 96.0\% \\ \hline
    \multicolumn{4}{c}{\textbf{Non context sensitive}} \\ \hline
    Others                  & 38 & 0  & 100\%  \\ 
    \bottomrule
  \end{tabular}
\end{table}


%% file: chapter/8_discussion.tex
\section{Discussion}

\subsection{Sharing concerns}
\bigbcp provide additional mechanisms for managing version history of non-native resources such as PDFs, images, and videos. It is important to note that native resources like Google Docs, Sheets, and Slides employ a different type of version history management. For non-native resources, users have the option to replace an existing file. For instance, when a user uploads a PDF file to their workspace and a file with the same name~(referred to as the old version) already exists, the user can choose to replace the old version with the new one.

While the replacement option offers convenience, it poses a potential vulnerability: the sharing attributes of the old version are also copied to the new version. This introduces a security risk, especially in relation to malicious \addon. These \addons can exploit the replace option by strategically creating a \textit{placeholder} file with crafted names, such as \textit{salary-report-2024.pdf}, and storing it in the victim's workspace. The \addons can then add the attacker as a collaborator for this resource. Later, when the user uploads the actual salary file and chooses to replace the placeholder file, the attacker gains automatic access to the salary report, even if the \addon is no longer installed on the user's workspace. This scenario highlights significant security implications that must be addressed to prevent unauthorized access facilitated by the replace option. 
This attack is practical and has been demonstrated in previous studies~\cite{nikiforakis2013bitsquatting} in other domains.

\subsection{Countermeasures}
In this section, we suggest countermeasures to help \bcp strengthen their security against vulnerabilities.

\noindent{\textbf{Fine-grained permission management for \addon:}} 
Currently, \bcp offer more stringent and fine-grained permission management for actual users. For example, spreadsheets allow for fine-grained permissions down to the row level. However, the inconsistency between strict user-level access control and the more coarse-grained add-on-level access control introduces numerous security vulnerabilities, as we have previously discussed. \bigbcp must enforce consistent and strict access control measures for \addons to protect sensitive resources.

\noindent{\textbf{Strict management of sharing privilege:}} 
\bigbcp maintain strict control over resource content, but their management of resource-sharing permissions is very loose and flat. 
We recommend that \bcp distribute only resource content-related permissions to collaborators, rather than resource management permissions. Besides, the sharing configuration should not be abused by \addons. Only allowing manual operation from the user side like the discussed API \texttt{transferOwnership()} would make the resource more secure.

%% file: chapter/9_limitations.tex
\subsection{Limitations} With all the findings discovered in our paper, it is still preliminary due to the following reasons. 
First, our work primarily focuses on the problem of three permission escalations.
Thus, \toolname is unable to detect unseen security vulnerabilities that do not meet our criteria. Second, for the rich context parameter of API renovation, we heavily rely on tutorial code snippets provided by \bcp. If not, we randomly choose a valid parameter value based on the \graph.  In the future, we can employ LLM with the \graph, feed API description as prompt to generate context-sensitive parameters that can work well even without tutorials.

%% file: chapter/10_related_work.tex
\section{Related Work}

\subsection{\textbf{Team workspace permission analysis.}} Previous studies mainly focus on the security and privacy of \bcp based on manual analysis. Wan et al. investigate the interaction between different user roles and resources by manual analysis. Their analysis reveals many security violations that lead to permission escalation. 
Our work builds on their findings and is the first to systematically scan for problematic APIs that may lead to permission escalation in \bcp.

Besides workspace, message systems like Slack and Microsoft Team also enable third-party applications to join as bots or delegators to invited channels and access the message history in team chat. Several studies~\cite{bcp_permission_analysis, hazard_team_chat, more_is_less, yan2025understanding, yan2024exploring} reveal that third-party apps in the chat system can hijack other apps, steal messages from channels that they are not invited, or perform malicious actions like merging GitHub pull requests without user's awareness. These security issues can lead to severe impacts considering the sensitive message history and resources in the team chat system. 

\subsection{\textbf{Third-party app security.}} The security of third-party applications has been widely studied across different domains like the \textbf{Browser}, \textbf{Android}, \textbf{Internet of things}~(IoT)  and \textbf{Message} systems.

\paragraph{\textbf{Browser side}}
AuthScan~\cite{bai2013authscan} identifies design flaws in the single-sign-on (SSO) web authentication protocol that result in seven security vulnerabilities and impacted millions of users. SSOScan~\cite{zhou2014ssoscan} extends this work by developing an automated vulnerability checker for applications using single-sign-on, revealing that many top-ranked websites are susceptible to SSO vulnerabilities. Some studies~\cite{trends_lesson_malicious_extension, pantelaios2020you} have demonstrated that browser extensions can be malicious and have created automated tools to detect malicious extensions.
Other studies~\cite{browser_ssp, browser_ase} have developed checkers for privacy policy~\cite{yan2024quality} declarations, identifying numerous violations.

\paragraph{\textbf{Android side}}
Wang et al.~\cite{android_intent_attack} conduct the first systematic study on mobile cross-origin risks and demonstrated that the absence of origin-based protection allows numerous attacks. Yang et al.~\cite{yang2022cross} and Zhang et al.~\cite{zhang2022identity} attempt to identify cross-mini-app security vulnerabilities in emerging mini-apps installed on WeChat or Snapchat.

\paragraph{\textbf{IoT side}}
Recently, researchers~\cite{security_privacy_ifttt, minTap, lazyTap} have expanded the scope of this topic to encompass IoT platforms such as IFTTT and voice assistance devices. Bastys et al.~\cite{ift_this_then_what} conduct the first analysis of IFTTT flows and discover numerous security vulnerabilities leading to confidential data leakage of third party applications. Mahadewa et al.~\cite{TAIFU_kulani} examine cross-app flows in IFTTT and identify many violations of privacy regulations. Additionally, SkillScanner~\cite{liao2023skillscanner} scrutinizes both the front-end and back-end code of Amazon Alexa skills, uncovering numerous privacy violations such as excessive data requests.

\paragraph{\textbf{Message system side}} Message systems like Slack and Microsoft Team also enable third-party applications to join as bots or delegators to invited channels and access the message history in team chat. Several studies~\cite{bcp_permission_analysis, hazard_team_chat, more_is_less} reveal that third-party apps in the chat system can hijack other apps, steal messages from channels that they are not invited, or perform malicious actions like merging GitHub pull requests without user's awareness. These security issues can lead to severe impacts considering the sensitive message history in the team chat system. 

In contrast to previous works, the distinctive nature of \bcp introduces new security vulnerabilities. For example, \bcp enable multi-user collaboration under various permission levels on the same resource, which introduces fresh access control risks such as permission escalation.

\subsection{\textbf{API analysis and testing.}} 
Unlike the commonly used REST APIs~\cite{hazard_team_chat} for web applications, which are flat, APIs in \bcp are more complex, featuring a well-structured design with interdependencies. Static analysis of such structured APIs~\cite{yan2024investigating} presents a challenge. Recent work such as IAceFinder~\cite{zhou2022uncovering} focuses on detecting access control inconsistencies between native (aka. C++) and Java contexts when accessing users' confidential data stored on mobile devices. IAceFinder and \toolname share similar goals and face similar challenges. They heavily depend on call graph analysis of Android libraries to generate test cases and identify security violations. In contrast to existing work, we must address API dependencies and parameter generation while considering the unique aspects of \bcp. To achieve this, we implemented the \toolname to manage dependencies and efficiently prune branches as needed, expediting our testing process.

\section{Conclusion}

We conduct a comprehensive study of the access control system within \bcp, focusing on resource management. We outline the two-level permission management framework for \addon to access sensitive resources and construct attack scenarios that lead to permission escalation, providing formal representations of these scenarios.

Utilizing the permission model established by \bcp, we developed and implemented an automated API testing tool called \toolname, which identifies APIs that diverge from the specified permission management protocols. \toolname successfully identified \riskyapi high-risk APIs and provided detailed  analyses. In response to the identified security vulnerabilities, we offer countermeasures to help \bcp mitigate these risks. We hope our analysis provides valuable insights into the security analysis of \bcp and encourages further research in this field. Our findings serve as a wake-up call for both \bcp and \addon developers.

\section*{Acknowledgement}
This research
has been partially supported by Australian Research Council Discovery Projects (DP230101196, DP240103068) and the Ministry of Education, Singapore under its Academic Research Fund Tier 3
(MOET32020-0003).